\documentclass[10pt]{article}
\usepackage[a4paper]{geometry}
\usepackage[T1]{fontenc}
\usepackage{algorithm,algorithmic}
\usepackage{hyperref}
\usepackage{authblk}
\usepackage{longtable,booktabs}
\usepackage[natbib,backend=biber]{biblatex}
\usepackage{amsmath}
\usepackage{graphicx,subcaption}
\usepackage[backgroundcolor=yellow]{todonotes}
\usepackage{lineno}
\usepackage{times}
\newcommand{\pkg}[1]{#1}

\title{Had enough of experts? Quantitative knowledge retrieval from large language models %
}
\date{}

\author[1,+,*]{David~Selby}
\author[1,2,+]{Yuichiro~Iwashita}
\author[1,+]{Kai~Spriestersbach}
\author[1]{Mohammad~Saad}
\author[3]{Dennis~Bappert}
\author[1]{Archana~Warrier}
\author[1]{Sumantrak~Mukherjee}
\author[1,2]{Koichi~Kise}
\author[1,4]{Sebastian~Vollmer}

\affil[1]{DFKI GmbH}
\affil[2]{Osaka Metropolitan University}
\affil[3]{Amazon Web Services}
\affil[4]{University of Kaiserslautern--Landau}

\affil[*]{\href{mailto:david.selby@dfki.de}{david.selby@dfki.de}}
\affil[+]{\textit{these authors contributed equally to this work}}

\begin{document}
\maketitle

\begin{abstract}\noindent
Large language models (LLMs) have been extensively studied for their abilities to generate convincing natural language sequences, however their utility for quantitative information retrieval is less well understood.
Here we explore the feasibility of LLMs as a mechanism for quantitative knowledge retrieval to aid two data analysis tasks: elicitation of prior distributions for Bayesian models and imputation of missing data.
We introduce a framework that leverages LLMs to enhance Bayesian workflows by eliciting expert-like prior knowledge and imputing missing data. Tested on diverse datasets, this approach can improve predictive accuracy and reduce data requirements, offering significant potential in healthcare, environmental science and engineering applications.
We discuss the implications and challenges of treating LLMs as `experts'.
\end{abstract}


\vspace{1ex}

\section{Introduction} \label{sec:intro} 

Automated solutions for life sciences, industrial and governmental processes demand large amounts of data, which are not always available or complete.
Small datasets are vulnerable to overfitting, weakening the validity, reliability and generalizability of statistical insights.
To overcome these limitations, analysts employ two approaches.
Data-based or empirical methods maximize information extraction, through imputation models, data augmentation and transfer learning; however, this is limited by the size, availability and representativeness of the training set.
Alternatively, one can exploit prior information, via knowledge graphs or expert-elicited Bayesian priors, allowing for sparser models and handling of missing values.
This approach is constrained by the difficulty, cost and myriad different methods of obtaining and eliciting subjective and heterogeneous opinions from experts, then translating them into a form amenable to quantitative analysis \citep{falconer_methods_2022}.

Large language models (LLMs) are generative models capable of producing natural language texts based on a given prompt or context.
LLMs such as GPT-4 have been used in various applications, such as chatbots, summarization and content creation.
In the quantitative sciences, LLMs have been applied to mostly qualitative tasks such as code completion, teaching of mathematical concepts \citep{wardat_chatgpt_2023} and offering advice on modelling workflows or explaining data preparation pipelines \citep{barberio_large_2023, hassani_role_2023}.
Some work has also applied LLMs to mathematical reasoning and symbolic logic \citep{he-yueya_solving_2023, orru_human-like_2023}.
When linked with certain application programming interfaces (APIs), or incorporated into a retrieval-augmented generation (RAG) tool, some LLM frameworks \citep[e.g.][]{ge_openagi_2023} are also capable of evaluating code, connecting to other data analysis tools or looking up supporting information \citep{nicholson_scite_2021, kamalloo_hagrid_2023}.
However, the capabilities of LLMs to retrieve accurate and reliable \emph{quantitative} information are less well-explored.

Here, we explore the use of LLMs for missing value imputation and for prior elicitation.
Data imputation is the process of replacing missing entries of a variable with substituted values based on empirical or informed estimates of the distribution \citep{donders_review_2006}.
Prior elicitation is eliciting knowledge from a domain expert to be converted into a prior distribution for use in a probabilistic Bayesian model \citep{mikkola_prior_2023}.

Can LLMs be considered `experts', having read a large sample of the scientific literature in their training corpora, and thus treated as an accessible interface to this knowledge?
We develop a prompting 
methodology to elicit prior distributions from LLMs, emulating real-world elicitation protocols.
LLM-elicited priors are compared with those from human experts, and the LLM `expertise' is quantitatively evaluated for several tasks.
We also present a zero-shot missing data imputation framework, based on LLMs playing `expert' roles derived from metadata such as the dataset description.
An empirical evaluation of imputation quality and impact on downstream tasks compares LLMs with baseline approaches on a diverse set of 50 real-world datasets.
Analysis code is available on \href{https://github.com/Selbosh/quantllm}{GitHub}.

\section{Related work} \label{sec:litreview}

Language models have been noted for their remarkable ability to act as unsupervised knowledge bases \citep{petroni_language_2019}.
\citet{noever_numeracy_2023, cheng_analyzing_2023} discuss the `emergent' numeracy skills of LLMs, from early models unable to perform simple addition to later versions able to compute correlations.
\citet{hopkins_can_2023} showed that repeated sampling from LLMs does not yield reasonable distributions of random numbers, making them poor data generators.
\citet{xiong_can_2023} also suggested LLMs tend to underestimate uncertainty.
It has been hypothesized that \emph{mode collapse} inhibits the diversity of outputs \citep{anonymous_understanding_2023}.
The design, adaptation and use of LLMs to assist data analysis is a broad topic.

Many LLM-based data science tools focus on tasks such as code generation for analysis scripts \citep{megahed_how_2023} 
or connection with external APIs \citep{ge_openagi_2023}.
A conversation with a chatbot can also offer generic advice on data science practices.
\citet{ahmad_retclean_2023} proposed a data cleaning model that combines a fine-tuned foundation model augmented with retrieval from a user-supplied data lake.
Here, however, we are interested in evaluating the intrinsic ability of an LLM to retrieve latent quantitative information directly; that is, not to perform mathematical operations on an input dataset nor to offer code or advice on how to do so, rather to offer educated numerical suggestions based on its large training corpus containing specialist technical knowledge.

There is some promise in converting data into natural language inputs for an LLM to perform preprocessing: \citet{narayan_can_2022} tested GPT-3 on entity matching, error detection and data imputation tasks, in zero-shot and few-shot settings.
Their approach serialized tabular data and tasks into natural language using manually tuned prompt templates.
\citet{vos_towards_2022} explored prefix tuning as an alternative to full fine tuning of an LLM for such tasks; whereas \citet{zhang_large_2023} compared GPT-3.5, GPT-4 and Vicuna-13B in a data preprocessing framework, later developing Jellyfish-13B, an open-source LLM fine-tuned specifically for data preprocessing~\citep{zhang_jellyfish_2023}.
\citet{li_table-gpt_2023}'s Table-GPT
describes a framework for fine-tuning language models on `table tasks', including finding and predicting missing values.
Separately, \citet{chen_gatgpt_2023} utilized fine tuning in tandem with a graph attention mechanism to impute spatiotemporal data.
\citet{nazir_chatgpt-based_2023} further explored the capability of ChatGPT in missing value imputation, focussing on imputation quality (see \autoref{sec:evalimpute}) in psychological and biological data.
Conversely, \citet{hayat_claim_2024} investigated the role of contextual information in ChatGPT's handling of missing values, but only for downstream LLM-based tasks.
An alternative approach to LLM-assisted data analysis involves using only the model's encoder to project natural language representations into a latent space, then performing anomaly detection on this embedding \citep{lopatecki_applying_2023, lopatecki_novel_2023}.
Tabular representation learning is an active area of research \citep{hollmann_tabpfn_2023}.

However, the level of `expertise' offered by pretrained LLMs on quantitative tasks across different domains has not yet been extensively studied, nor the effect of LLM imputations on performance in downstream supervised learning tasks.

Prior distributions are just one form of knowledge elicited from domain experts; others include feature engineering, model explanations and labelling heuristics, but in each case the process of elicitation typically involves interviews, written correspondence or interaction with a custom app \citep{kerrigan_survey_2021}.
A good expert-elicited prior distribution can help a statistical model effectively represent the data generating process, although due to various practical, technical and societal factors, prior elicitation is not yet widespread practice.
A lack of standardized software means there is no way for an analyst using, e.g. Stan, to initiate an elicitation exercise for a specific model \citep{mikkola_prior_2023}.

LLM-\emph{driven} elicitation \citep{li_eliciting_2023} uses an LLM to assist elicitation from human experts, making the process interactive.
In engineering, LLMs have been employed in generating (and responding to) requirements elicitation surveys \citep{white_chatgpt_2023, ronanki_investigating_2023, gorer_generating_2023}.
Natural language processing is already extensively used to extract quantitative information from large texts to aid quantitative research \citep[see, e.g.][]{olivetti_data-driven_2020} and to curate structural information for causal models \citep{zhang_leveraging_2024}.
Prior distributions can be elicited from literature via systematic reviews \citep{rietbergen_incorporation_2011, van_de_schoot_bayesian_2018, linde_data-driven_2023}.
A meta-analytic-predictive prior uses historical data to reduce the required sample size in clinical trials \citep{weber_applying_2021}.
To our knowledge, direct elicitation of parametric priors from a `domain expert' LLM has not yet been explored.
\citet{gouk_automated_2024} generated pseudodata as an indirect prior elicitation approach for Bayesian logistic regression; by contrast, in this paper we attempt to elicit the distributional parameters directly.
Our work has already been extended by \citet{capstick_using_2024}, who elicit multiple Gaussian distributions as priors to obtain a mixture model.



Several elicitation protocols have been developed to mitigate cognitive biases and combine the judgements of multiple experts \citep{ohagan_expert_2019}.
The Sheffield Elicitation Framework \citep[\textsc{Shelf};][]{gosling_shelf_2018} describes a collection of methods for eliciting a distribution based on aggregated opinions of multiple experts, through group discussion guided by a facilitator.
Following a primer in probability and statistics, the protocol includes various ways of eliciting a univariate distribution, such as the `roulette method', where participants assign chips to bins to form a histogram.
Alternatively, the quartile method \citep[or `Sheffield method';][]{european_food_safety_authority_guidance_2014} uses a series of questions to elicit quantiles of a distribution.
Cooke's method \citep{cooke_experts_1991} pools the distributions of multiple experts, weighted according to their respective `calibration' (accuracy) 
 and `information' (uncertainty). 
The Delphi method uses the quartile method, iteratively refined over successive rounds using anonymized feedback from other participants.
In this paper, however, we consider only single-agent LLMs with a zero-shot approach.


\section{Methods} \label{sec:methods}

\subsection{Overview}

We describe our framework for generating and assessing `expert' prior distributions and imputed missing values for tabular datasets. Evaluation criteria include prior informativeness, calibration and imputation quality, benchmarked against statistical baselines.
Since an empirical evaluation of an LLM's `real-world' knowledge precludes purely abstract simulation-based studies, we carefully sample real datasets spanning a diversity of scientific and technical disciplines.
Prior elicitation is tested on domain-specific knowledge tasks, while imputation is evaluated using classification benchmark datasets. These experiments assess both prior elicitation and imputation across multiple domains.

\subsection{Eliciting priors and imputed values from LLMs}
\label{sec:prompting} 

Impersonating a human domain expert can improve an LLM's performance at related tasks \citep{salewski_-context_2023}.
Nevertheless, in response to scientific questions, especially on potentially sensitive topics, such as healthcare advice, language models often prevaricate \citep{lautrup_heart-to-heart_2023}.
An LLM elicitation system should therefore not only prompt the model to roleplay an expert, but also carefully specify the task to ensure contextually relevant information is returned in the appropriate format.

Our \emph{expert prompt initialization} module is a system prompt defining a suitable expert role for the model to imitate.
For efficiency, the LLM itself is used to generate these descriptions, once per task, of the form ``You are a...''.
To avoid the model offering verbose, generic or prevaricating advice about prior elicitation (such as suggesting R or Python code or advising the user to consult a real expert), the \emph{task specification} module insists that the model follows a particular elicitation protocol followed by returning a parametric prior distribution, e.g. ``\texttt{Beta(1, 1)}'', or an imputed number in a format that can be parsed programmatically.

Once an appropriate expert role is defined, we structure the input data in a format suitable for LLM processing.
Whereas some authors \citep[e.g.][]{zhang_jellyfish_2023, li_table-gpt_2023}
pass data to the LLM in a tabular format, we serialize it to a natural language form using our \emph{data serialization} module \citep[as in][]{nazir_chatgpt-based_2023}, described in Algorithm~\ref{alg:ds}, allowing the system to emulate interaction with a human expert.
Note, however, that in the case of prior elicitation, no observed data are passed to the model at all.
During imputation, the only data fed to the model are the values of other features from the same row---thus the LLM may not perform symbolic reasoning or otherwise average over other samples in the dataset.
In line with earlier work \citep{narayan_can_2022, vos_towards_2022, nazir_chatgpt-based_2023} we serialize tabular data using a simple template structure `\texttt{the \{variable\} is \{value\}}': for example a row-vector of data $(37, M, 120)$ with column names `Age', `Sex' and `Blood Pressure' would become the sentence `\texttt{The Age is 37. The Sex is M. The Blood Pressure is 120}'.
Though one might be tempted to add units or expand abbreviations, this conversion is necessarily deterministic to avoid data corruption.
Missing values (that are not to be imputed) are simply omitted from the prompt.

To mitigate stochasticity and redundant computation, our framework assumes a temperature setting of zero.
Further details are given in the appendix and our code is available on \href{https://github.com/Selbosh/quantllm}{GitHub}.

To ensure robustness in evaluation, we conduct empirical tests using datasets spanning multiple fields. We compare LLM priors with those from human experts and simple statistical models fitted to real-world data. The effectiveness of priors is further assessed by examining their predictive consistency across domains.

\subsection{Evaluating expert priors}

What makes a good prior?
Bayesian statistics involves decisionmaking based on a posterior distribution, 
\[p(\theta|D) \propto \pi(\theta) \prod_{i=1}^np(x_i|\theta),\]
where $\pi(\theta)$ denotes the prior distribution and $D$ the observed data.

The definition of a `good' prior distribution---like Bayesian statistics itself---is subjective, depending on the analyst's understanding of the purpose of expert-elicited information.
No standard benchmark exists for expert-elicited prior distributions; a prior is a function of the expert and the elicitation method, as well as of the predictive task \citep{gelman_prior_2017}.
One purpose of prior information is to reduce amount of data needed. Another is to treat expert knowledge and observed data as complementary sources of information about a natural process.
Any statistical model is at least slightly misspecified, but a prior can still be \emph{informative}, \emph{well-calibrated} and \emph{useful} \citep[see][]{williams_comparison_2021}.
An informative prior is different from a non-informative or default prior, i.e.\ it is not too vague.
Well-calibrated or `realistic' priors should align with those from human experts or be otherwise externally verifiable.
`Useful' means superior posterior predictive performance on a downstream task, improving expected utility over reference priors.
Here we consider informativeness and calibration.

A measurement of the informativeness of a prior distribution is the prior effective sample size \citep{morita_determining_2008, neuenschwander_predictively_2020}.
This is not data dependent and does not measure improvement on downstream tasks, but rather how many data points one would need to get similar peakiness/curvature around the posterior mode.
The heuristic prior effective sample size for $\text{Beta}(\alpha, \beta)$ is $\text{ESS}=\alpha+\beta$ \citep{morita_determining_2008}, which measures the concentration of the prior and the amount of data needed to shift the posterior if the prior were misspecified.

For measuring data--prior calibration, we can use the Bayesian log posterior predictive density, or lppd \parencite{mcelreath_statistical_2016}---also called log loss---or the continuous ranked probability score (CRPS), a proper scoring rule used in weather forecasting \parencite{gneiting_strictly_2007}.
We can estimate either metric using the posterior predictive distribution
\( p(\mathbf{x}'|D)=\mathbf{E}_{p(\boldsymbol\theta|D)}[p(\mathbf{x}'|\boldsymbol\theta)] \)
on held-out data. 
\citet{wilde_foundations_2021} describe a similar approach quantifying utility of synthetic data.

\subsection{Prior elicitation experiments} \label{sec:experiment-prior}

Our exploration of LLM-elicited priors includes three experiments: (1) a qualitative comparison with human-elicited priors; (2) estimating expert confidence or informativeness across different tasks; (3) a quantitative evaluation of calibration on real-world datasets. A fourth experiment is given in the appendix.

Absent an open benchmark or database of expert-elicited priors, we select a recent work from the literature that describes an elicitation procedure and reported the resulting distributions.
\citet{stefan_expert_2022} interviewed six psychology researchers about typical small-to-medium effect sizes 
and Pearson correlations 
in their respective specialisms, using the histogram method.
In our first experiment, using similar question wording, we elicited prior distributions from LLMs prompted to simulate an expert, conference of experts \citep{phillips_decision_1991} 
or non-expert, with and without mentioning the \textsc{Shelf} protocol.
This experiment is a qualitative comparison of how LLMs behave when emulating a published example of a prior elicitation exercise with published question wording and results.

For our second experiment,
we prompted GPT~3.5 to formulate 25 tasks that might call for expert elicitation in the fields of healthcare, economics, technology, environmental science, marketing and education.
Tasks correspond to proportions or probabilities following a $\operatorname{Beta}(\alpha, \beta)$ distribution.
These scenarios were then used to gauge general levels of confidence of elicited distributions from different LLMs, using the prior effective sample size metric for this distribution: $\text{ESS}=\alpha+\beta$ \citep{morita_determining_2008}.

Thirdly, we attempted to quantify the realism of priors and any prior--data conflict, by estimating how many samples the LLM prior offers for an analyst who has not yet collected any data.
Specifically, we compared LLM priors with simple models fitted to readily-available meteorological data.
Priors were elicited from LLMs for the typical daily temperature ($^\circ$C) and precipitation (mm) in December for 25 small and large cities around the world.
These distributions were then compared with historical weather data from the \texttt{openmeteo} API.
By investigating different continents and varying sizes of settlements, the goal was to identify any systematic biases that might emerge from LLMs' respective training corpora.
Exploring both temperature and precipitation allow us to observe behaviour with symmetrical and skewed distributions, respectively.

We compared the prior predictive distribution to a supervised learning model with conjugate posterior \citep{gressmann_probabilistic_2019}: a normal-inverse-gamma model for temperature and a gamma-exponential for precipitation. 
We ask: how many samples on average would a frequentist model need to achieve the same or better log-loss (or CRPS or MSE) than the prior predictive distribution?
We split the data in half for testing and repeatedly sample up to $\frac{1}{3}$ for training from the remaining half. An alternative comparison would be of a posterior predictive based on data and a baseline prior, however choosing such a baseline is difficult.
Unlike the ($\alpha+\beta$) effective sample size heuristic, this data-dependent approach quantifies prior--data conflict.

\subsection{Evaluating missing data imputation} \label{sec:evalimpute}

What makes a good imputed value?
\citet{jager_benchmark_2021} describe two principles of benchmarking imputation methods: \emph{imputation quality} and \emph{downstream evaluation}.
Imputation quality---or upstream performance---measures the extent to which an imputation method can accurately recover artificially missing values.
Downstream evaluation measures the predictive performance of a supervised learning model on the imputed dataset.
Imputation quality for continuous features can be calculated using the root mean square error,
$\text{RMSE} = \sqrt{\frac1n \sum_{i=1}^n (x_i - \hat{x}_i)^2}$,
where $x_i$ represents the original discarded value and $\hat{x}_i$ the output of imputation, whereas for categorical features, imputation quality can be calculated via the $F_1$ score,
$F_1 = 2(\text{recall}^{-1} + \text{precision}^{-1})^{-1}$, the harmonic mean of precision and recall.
As RMSE is unbounded, inter-dataset comparison is made possible by using a normalised version: $\text{NRMSE} = \text{RMSE}/(\max x - \min x)$.
Downstream performance is then determined from the relative improvement in predictive performance of models trained on imputed (\textit{vs.} complete-case) data.

\subsection{Imputation experiments} \label{sec:experiment-impute}

Our imputation experiments expand on previous work by exploring a wider range of applications.
The OpenML-CC18 Curated Classification benchmark \citep{bischl_openml_2017} comprises 72 classification datasets, based on real-world classification tasks, in a variety of domains from credit scoring to biology, medicine and marketing.
Use of this collection ensures our experiments cover a wider set of domains than previous work on LLM data imputation; meanwhile a pre-specified benchmark mitigates the risk of `cherry picking'. 
Datasets in CC18 have sample sizes ($n$) from about 500 to tens of thousands, with numbers of features ($p$) ranging from 5 to 3073.
Though all ostensibly provided in dense tabular format, some are actually drawn from other modes; for example, the MNIST and CIFAR-10 imaging datasets are represented as wide tables, with columns corresponding to individual pixels.
It is fair to assume that any human-like expert is unlikely to make particularly informed imputations about such features.
Further dataset details are given in Appendix~\ref{appendix:openml-cc18}.

To evaluate the imputation quality, we used all 64 of the datasets that did not already contain missing values.
We then split each of the datasets into training and test sets respectively comprising 80\% and 20\% of the samples.
For each dataset, we artificially generated missing values based on the missing at random (MAR) missingness pattern, where the probability of a value being missing depends only on the observed values, using the Python package Jenga~\citep{schelter_jenga_2021}.
The number of features affected by missingness was set to $\min(p, 3)$, and the number of missing values set to 40 for the training set and 10 for the test set.


Building on \citet{jager_benchmark_2021} and \citet{nazir_chatgpt-based_2023}, our LLM data imputer is compared with three data-driven approaches: mean and mode imputation (for continuous and categorical features, respectively), $k$-nearest neighbours ($k$-NN) imputation (the respective mean/mode of the $k$ nearest samples) and random forest imputation.
Mean/mode imputation served as the primary baseline.
The LLM-based data imputer was powered by Llama 2 13B Chat, Llama 2 70B Chat \citep{touvron_llama_2023}, Mistral 7B Instruct \citep{jiang_mistral_2023} and Mixtral 8x7B Instruct \citep{jiang_mixtral_2024}, each evaluated separately.

Upstream performance was calculated as the average imputation quality across selected features.
Downstream evaluation used a random forest classifier (a \texttt{RandomForestClassifier} from \pkg{scikit-learn}~1.3.2 with default hyperparameters) trained on the imputed training data and evaluated on the held-out test sets.



\section{Results}

\subsection{Prior elicitation}
While variations in prompting methodology---asking LLMs to roleplay as experts or non-experts---had some effect on the elicited priors, greater differences were observed between models than between tasks.
Variation between human experts is mirrored by variation between LLMs, but there was no obvious pattern to explain performance at particular categories of tasks.
Repeated queries with the same prompts offered mostly consistent responses, which can attributed to the low temperature setting.

\begin{figure}[p]
\centering
\includegraphics[width=\linewidth]{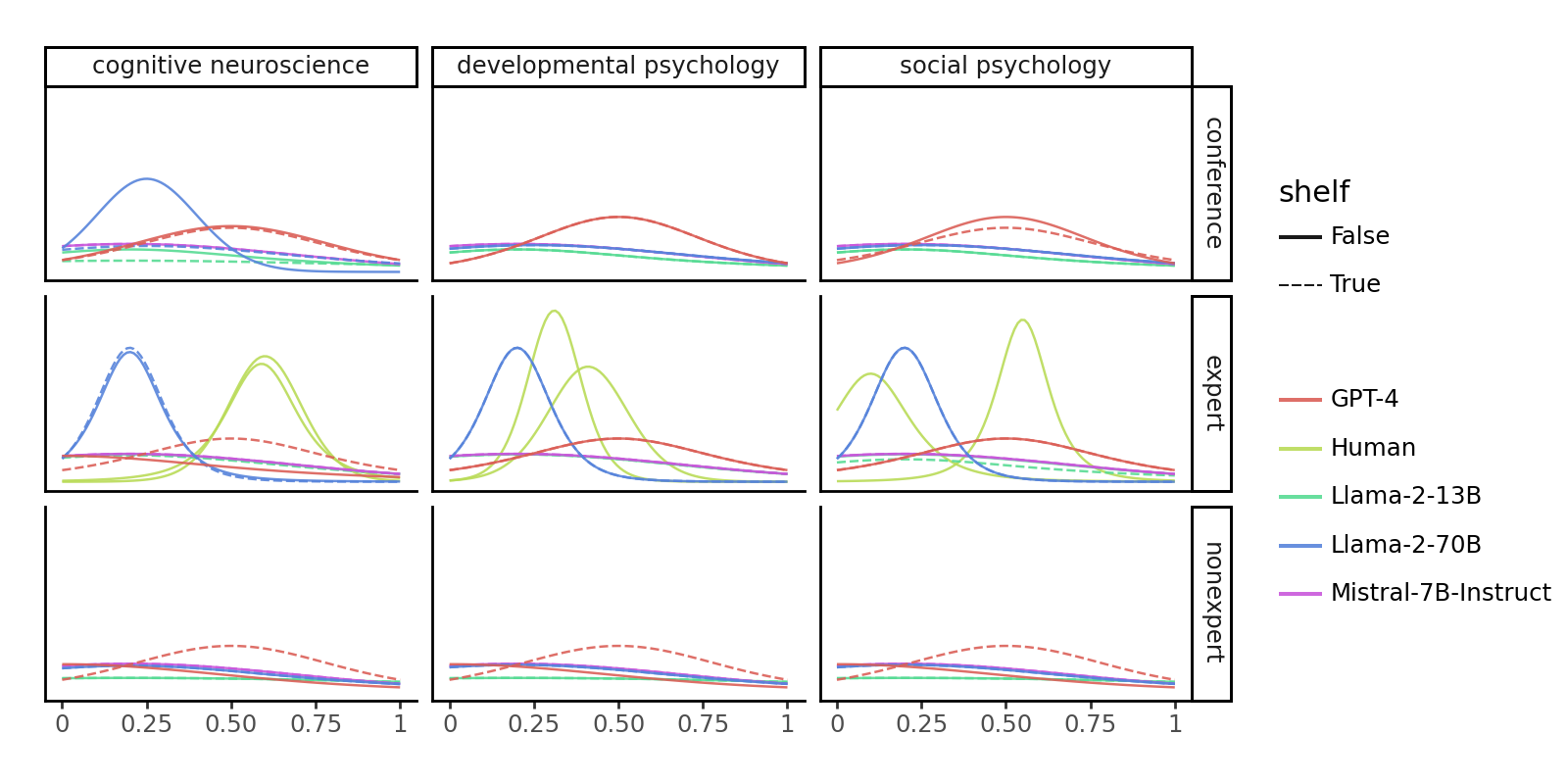}
\includegraphics[width=\linewidth]{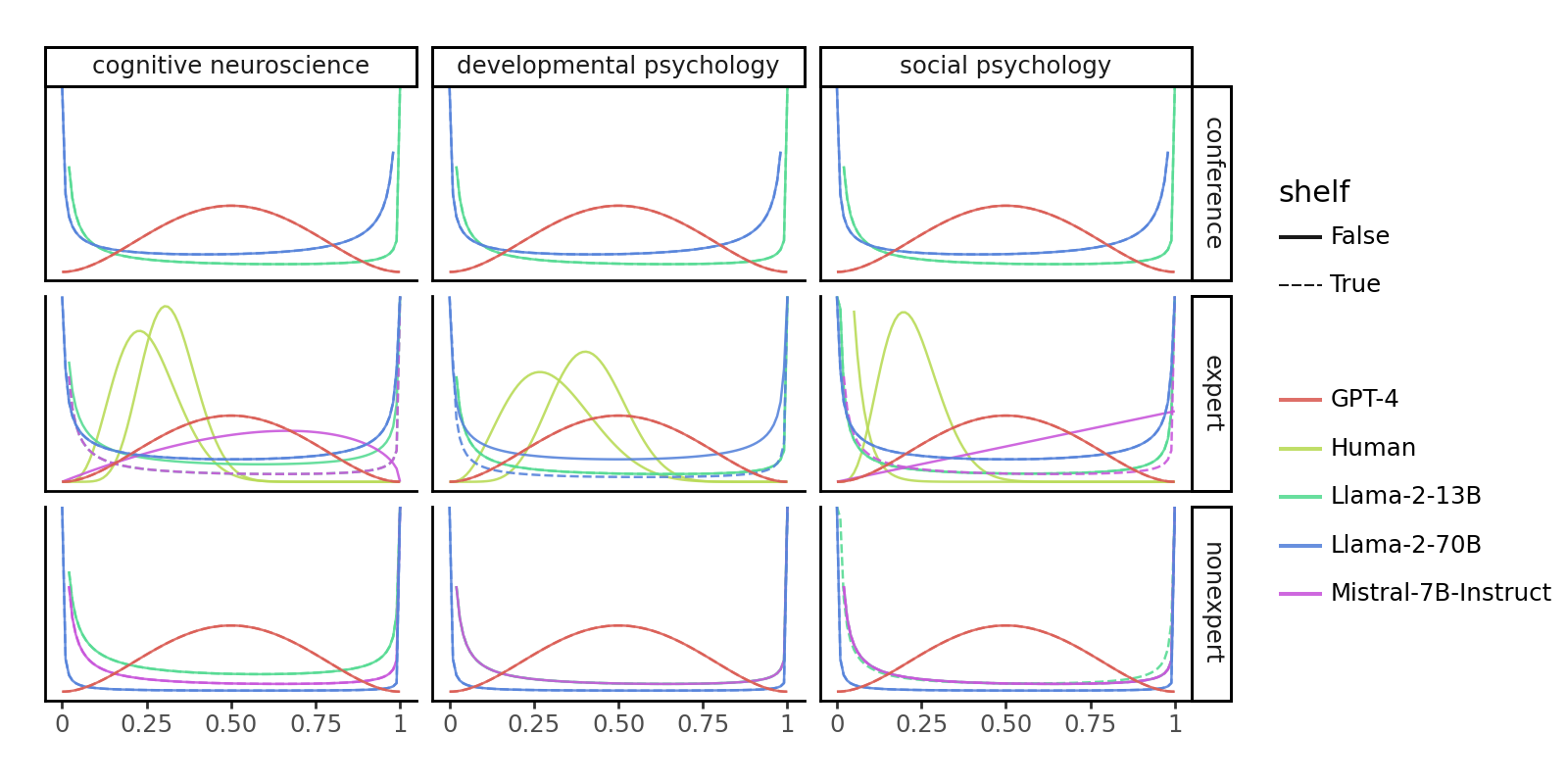}
\caption{Priors for Cohen's $\delta$ (top) and Pearson correlations (bottom) elicited from LLM and human experts in psychology. Dashed lines denote a \textsc{Shelf}-like elicitation protocol}
\label{fig:stefan}
\end{figure}

\autoref{fig:stefan} compares the priors elicited by \citep{stefan_expert_2022} from human experts with those we elicited from LLM counterparts in the fields of social and developmental psychology and cognitive neuroscience.
Roleplaying as experts in different sub-fields did not have a noticeable effect on the priors.
LLM priors for Cohen's $\delta$ were mostly centred around small effect sizes of 0.2--0.25, except GPT-4, which offered distributions around $\delta=0.5$.
Mistral-7B-Instruct invariably gave $t$ distributions with $\nu=30$ (Llama-70B-Chat-Q5: $\nu=5$); other models appeared to grow more conservative (smaller $\nu$; more leptokurtic distributions) if asked to roleplay as an expert, simulate a decision conference or employ the \textsc{Shelf} protocol.
Pearson correlation beta priors from LLMs apparently had little in common with those from real experts: GPT-4 provides a symmetric unimodal distribution whereas other models offer a right-skewed `bathtub' distribution.
Like real experts, different LLMs offered different opinions to each other.

\begin{figure}[p]
\centering
\includegraphics[width=0.9\linewidth]{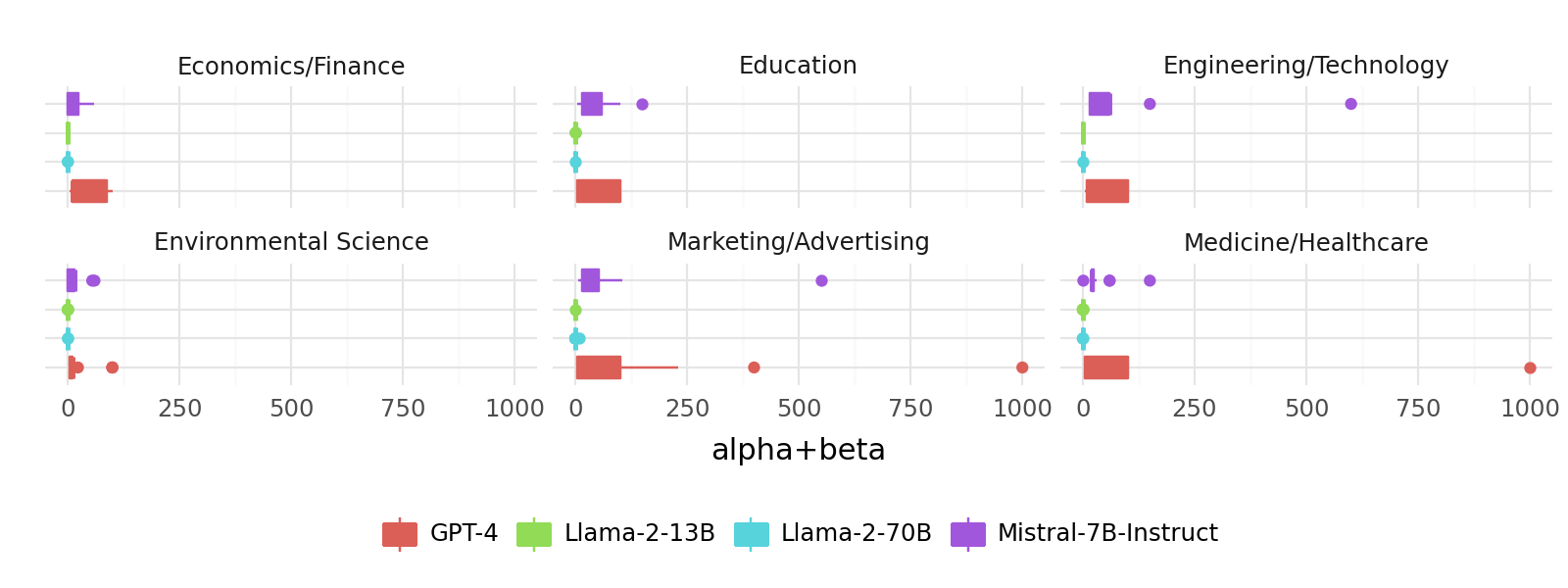}
\caption{Distribution of prior effective sample size ($\alpha+\beta$) for beta priors on various tasks. Outliers are omitted}
\label{fig:task-beta}
\end{figure}

From the second experiment,
\autoref{fig:task-beta} shows the prior effective sample size---corresponding to model confidence or informativeness---for priors on different tasks.
Llama-based models appear to give more conservative priors, whereas GPT is consistently more informative.
Mistral 7B Instruct occasionally offered extremely high values $\alpha\geq1000$, but, generally speaking, there was no clear and systematic difference between domains.

\begin{figure}[p]
\centering
    \begin{subfigure}{0.49\linewidth}
    \includegraphics[width=\linewidth]{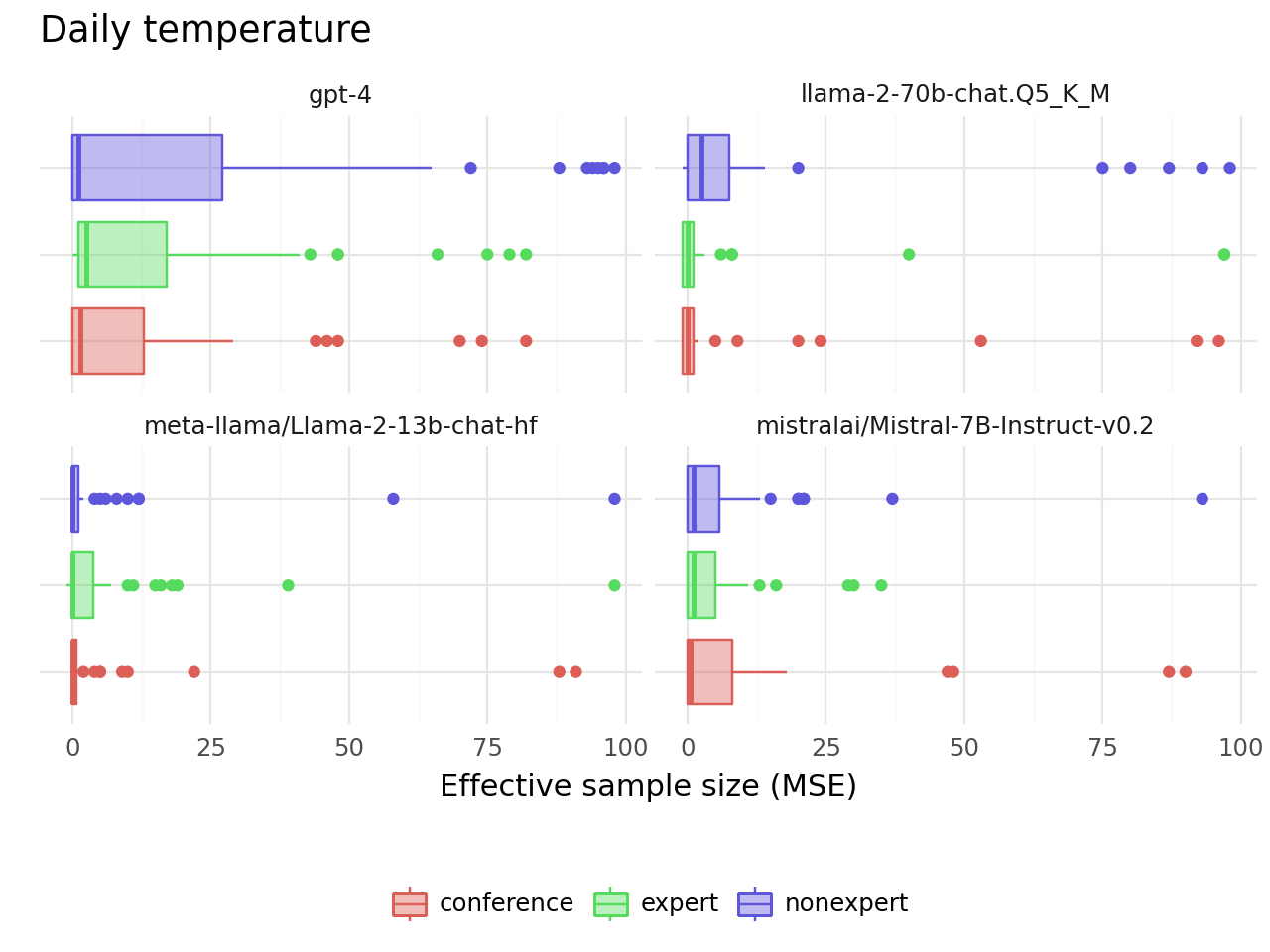}
    \caption{Daily temperature ($^\circ$C)}
    \end{subfigure}
    \hfill
    \begin{subfigure}{0.49\linewidth}
    \includegraphics[width=\linewidth]{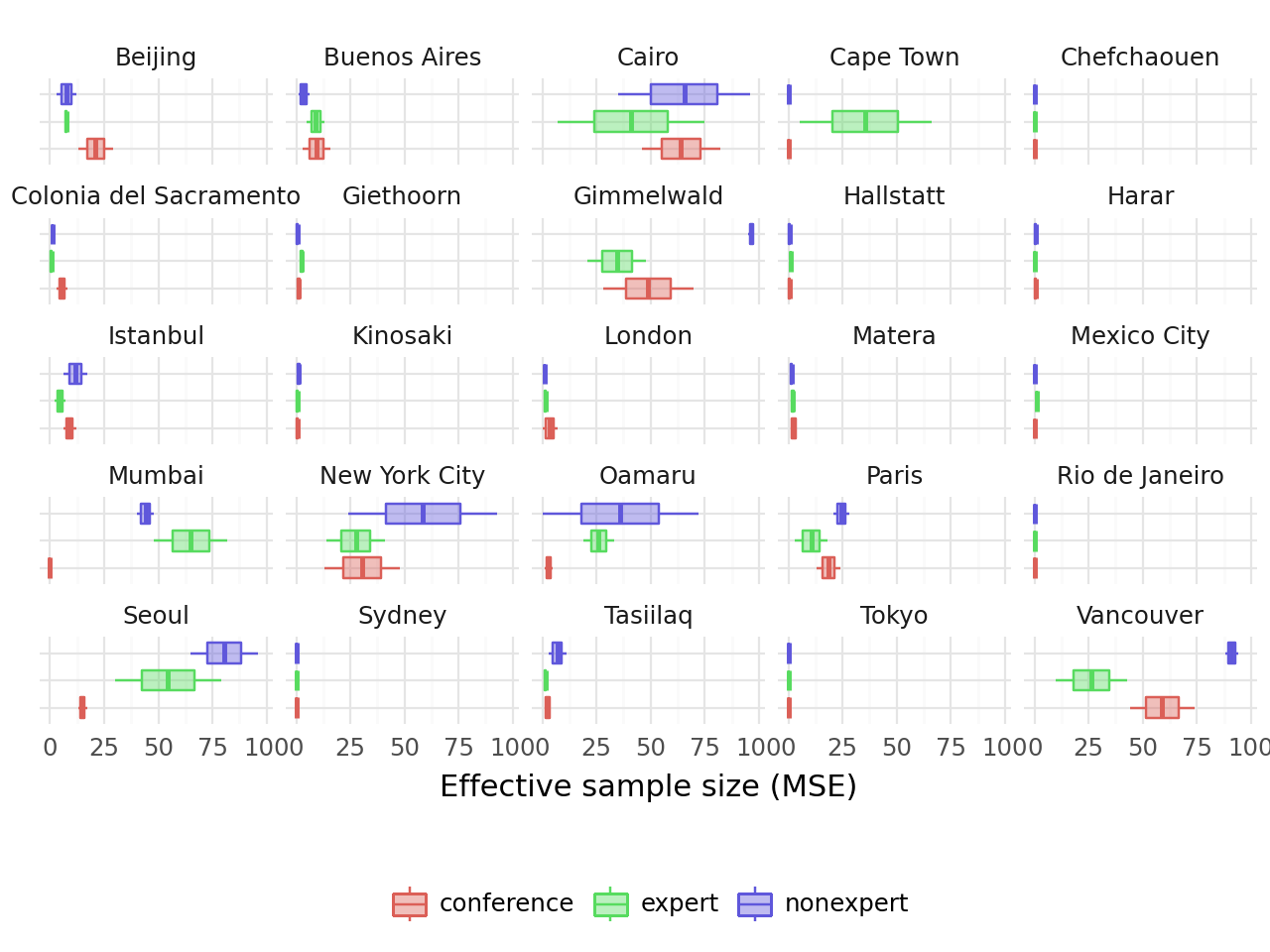}
    \caption{Daily temperature ($^\circ$C)}
    \end{subfigure}
    \begin{subfigure}{0.49\linewidth}
    \includegraphics[width=\linewidth]{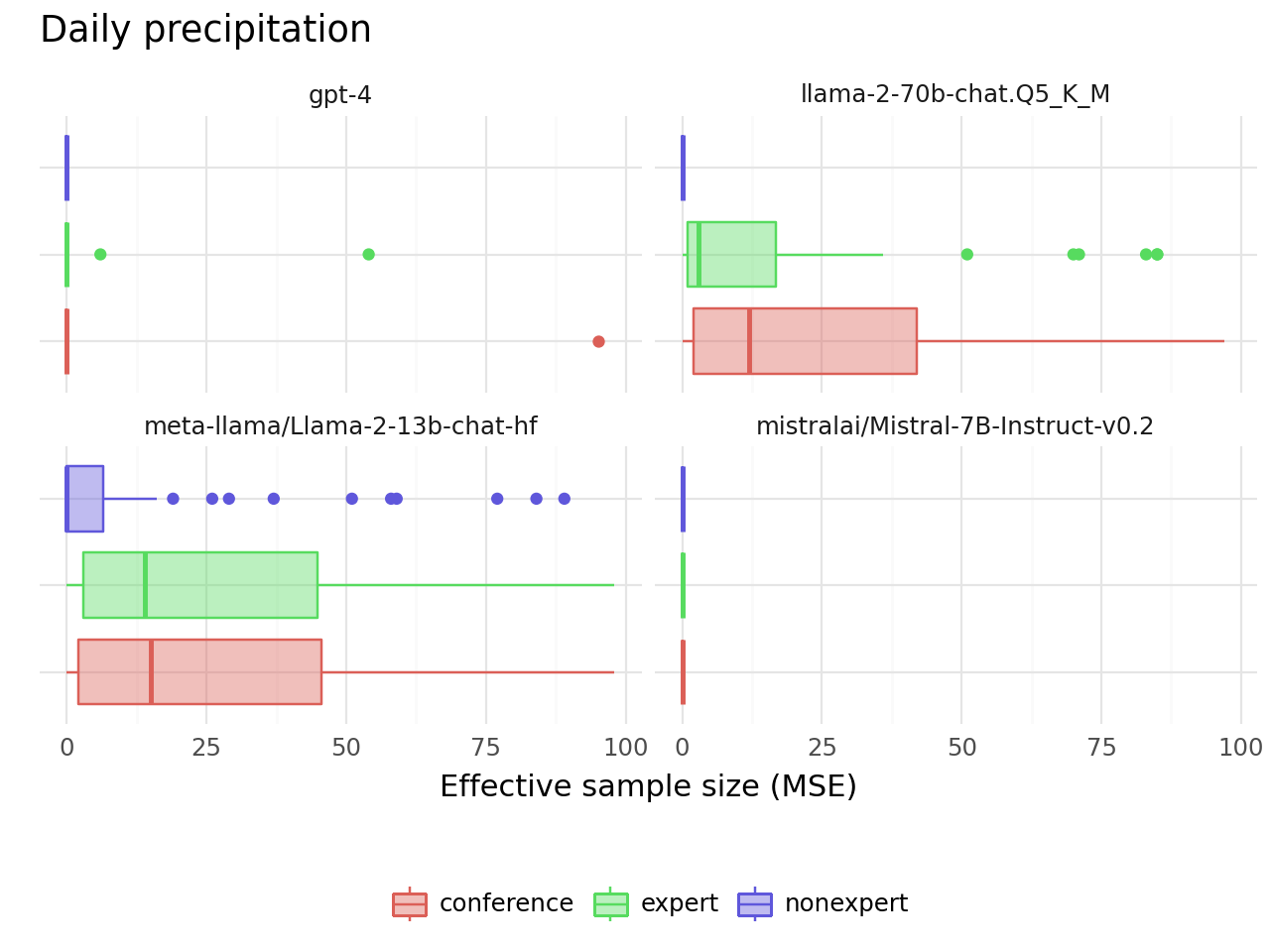}
    \caption{Daily precipitation (mm)}
    \end{subfigure}
    \hfill
    \begin{subfigure}{0.49\linewidth}
    \includegraphics[width=\linewidth]{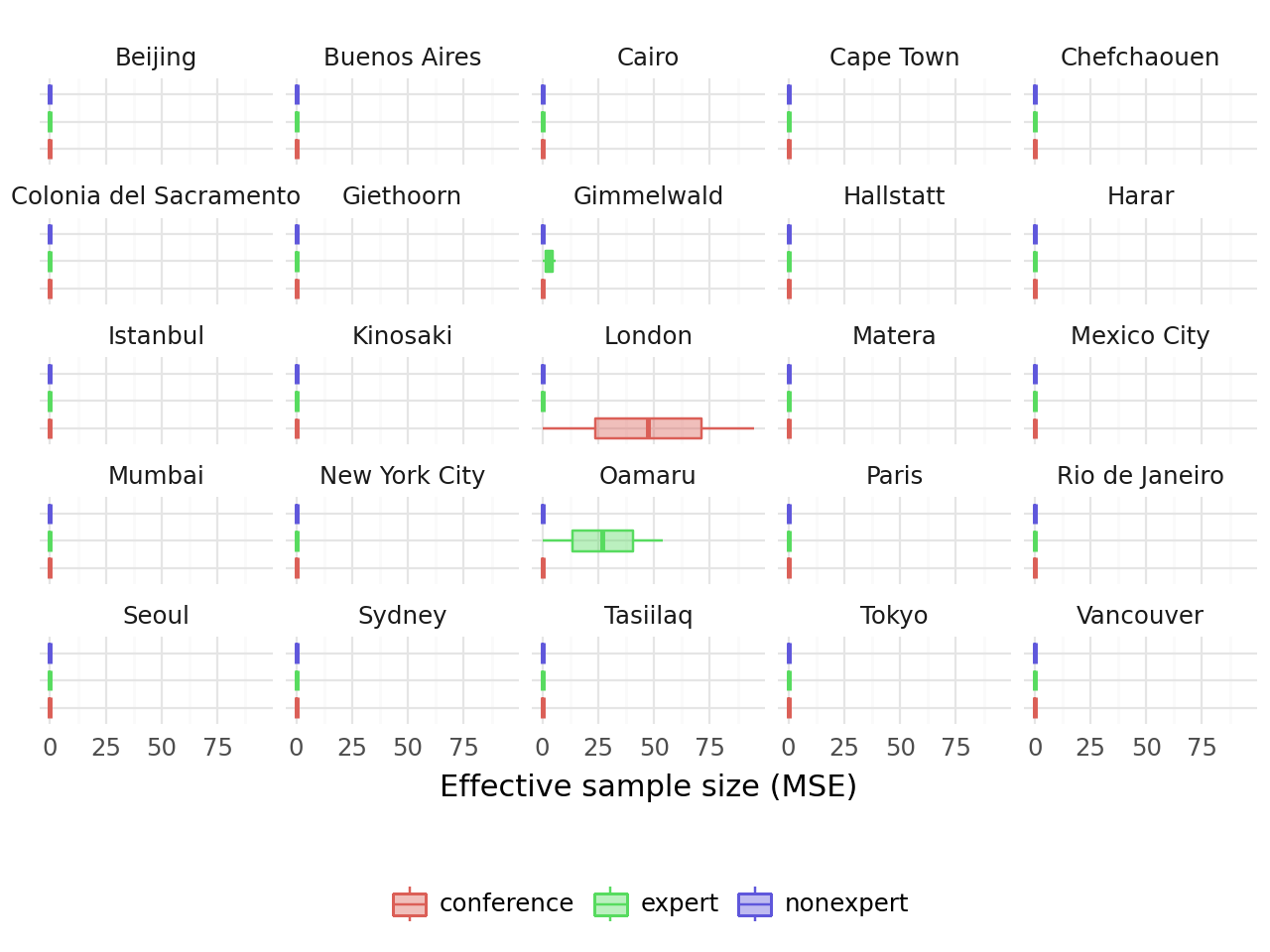}
    \caption{Daily precipiation (mm)}
    \end{subfigure}
    \caption{
    Benefit of LLM priors for weather forecasting: number of observations needed for a frequentist model to achieve better MSE than the prior predictive distribution \label{fig:weather}
    }
\end{figure}

Our third task was a data-driven evaluation on LLM meteorological estimates.
Figure~\ref{fig:weather} shows the data-dependent effective sample sizes from the elicited prior predictive distribution.
We measure the effective increase in observations (starting from zero samples) for a frequentist model to obtain better mean squared error (MSE) than the elicited prior predictive distribution.
The effective sample size (ESS) is the number of samples needed by the frequentist model to outperform the prior predictive model.
In many cases, the prior is in conflict with the data and so $\text{ESS}\rightarrow 0$ (or, strictly speaking, 2, the minimum data points needed to compute an empirical standard deviation).
As might be expected, there was a slight (though not necessarily overwhelming) bias in favour of larger, more populous cities, which would be better represented in the LLMs' training corpora.
However, a strong trend of better predictions for English-speaking countries (irrespective of size) was not apparent.

\subsection{Missing value imputation}
\label{sec:imputation-results}
Data imputation tasks revealed greater variation between domains, however LLMs did not consistently outperform more traditional imputation methods.

\begin{figure}[p]
\centering
\includegraphics[width=0.45\linewidth]{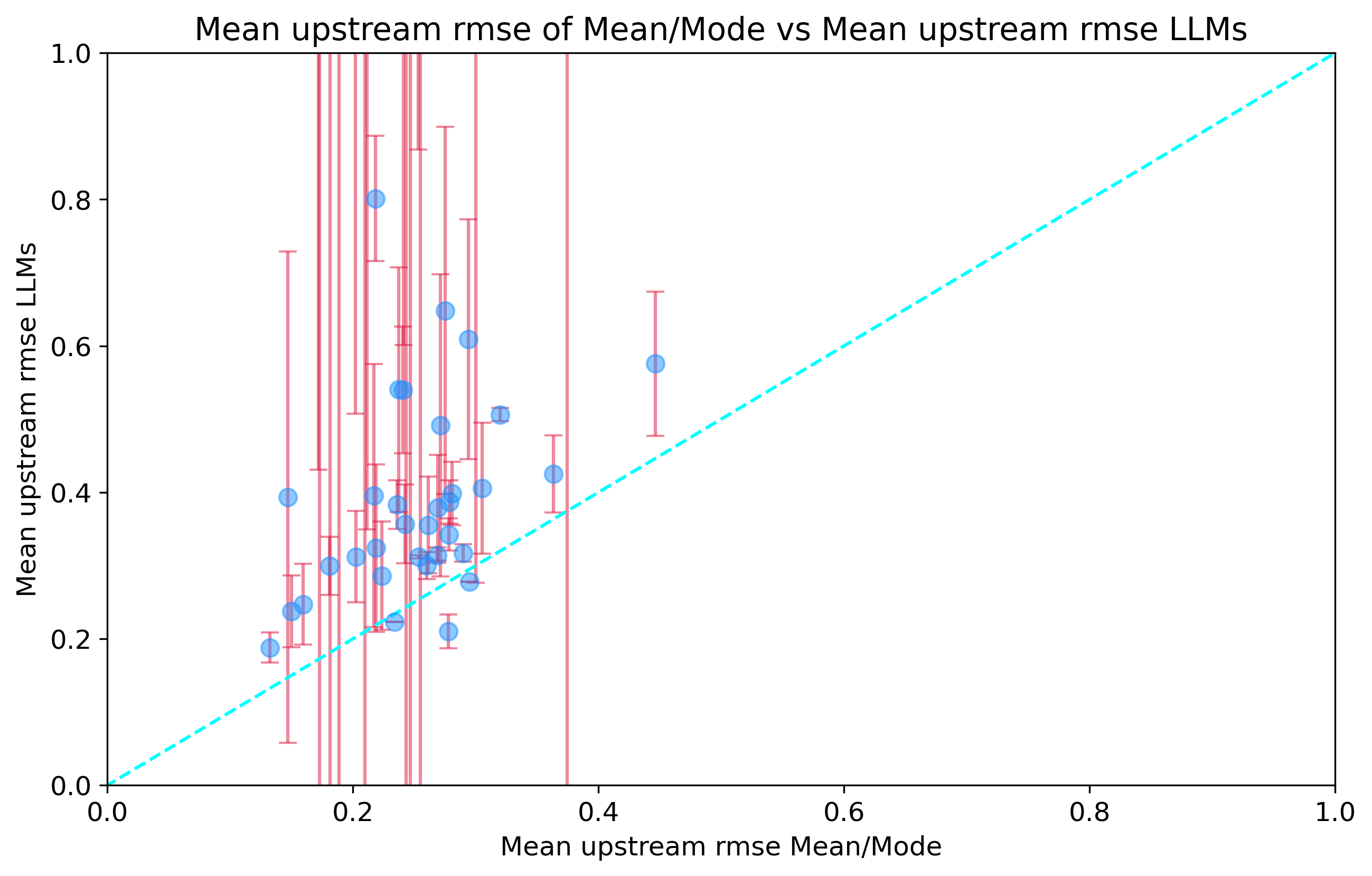}
\includegraphics[width=0.45\linewidth]{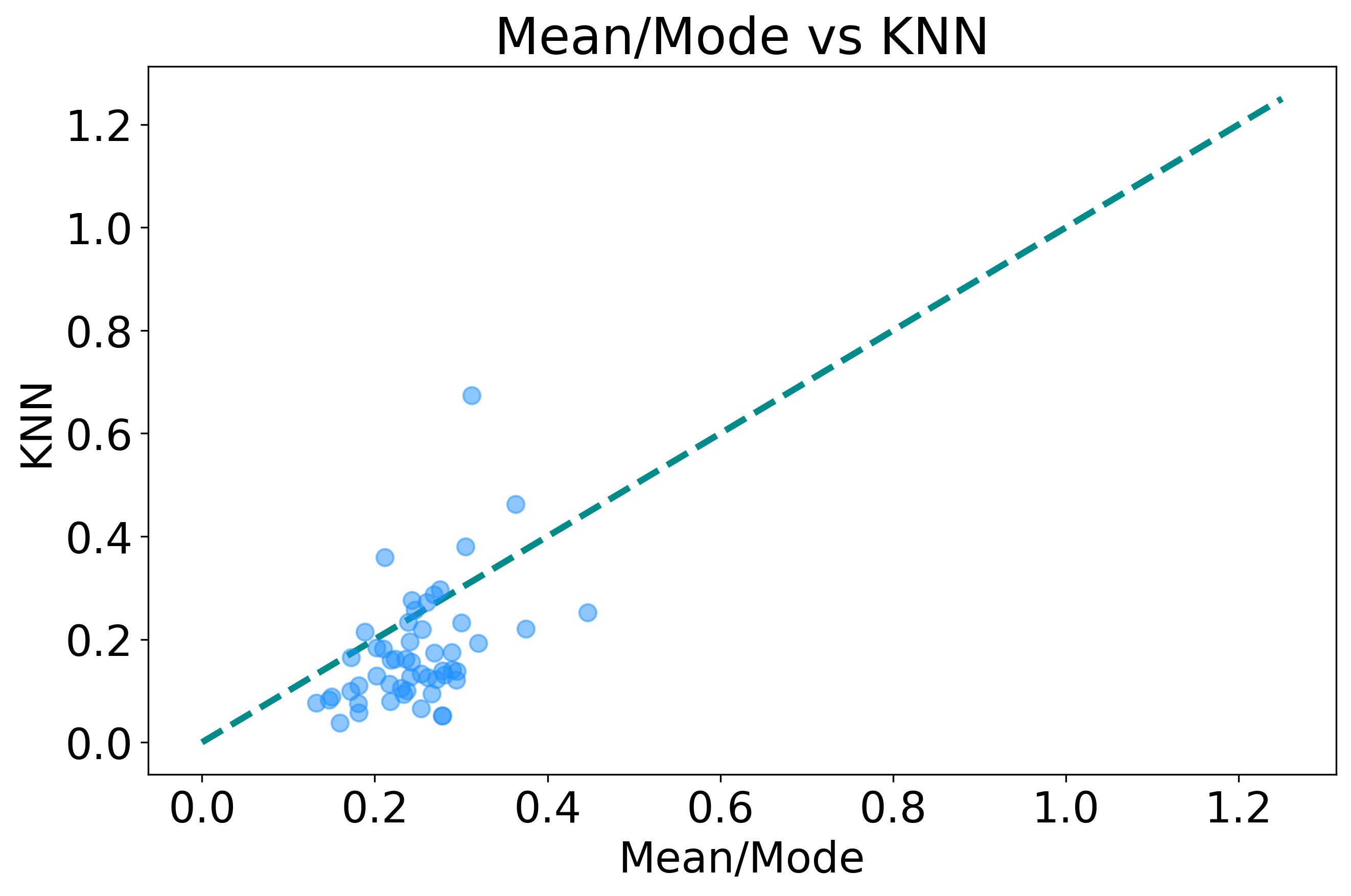}
\includegraphics[width=0.45\linewidth]{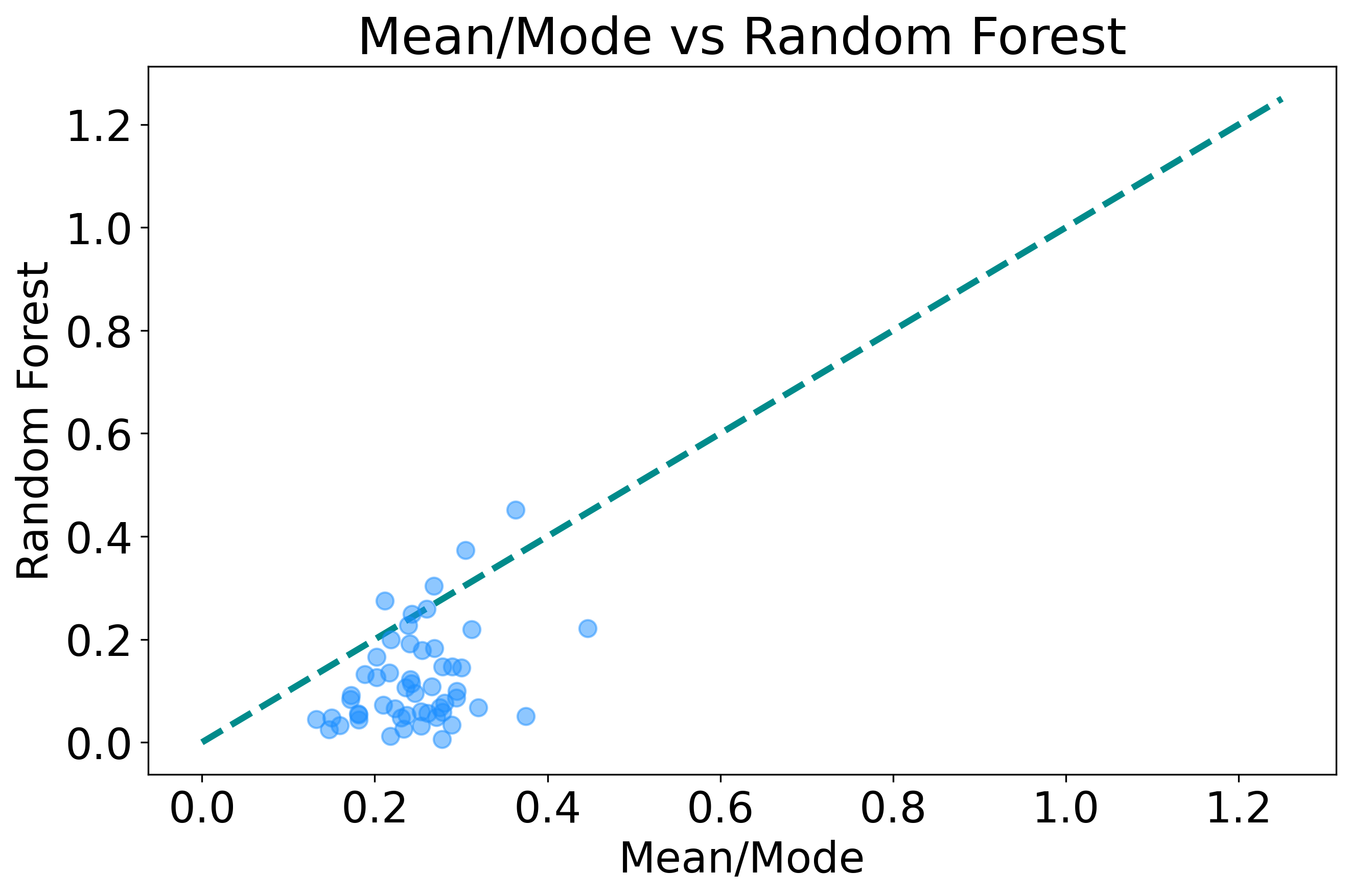}
\caption{Upstream RMSE performance across LLMs, KNN and random forest}
\label{fig:upstream-rmse}
\vskip -0.2in
\end{figure}

\begin{figure}[p]
\centering
\includegraphics[width=0.6\linewidth]{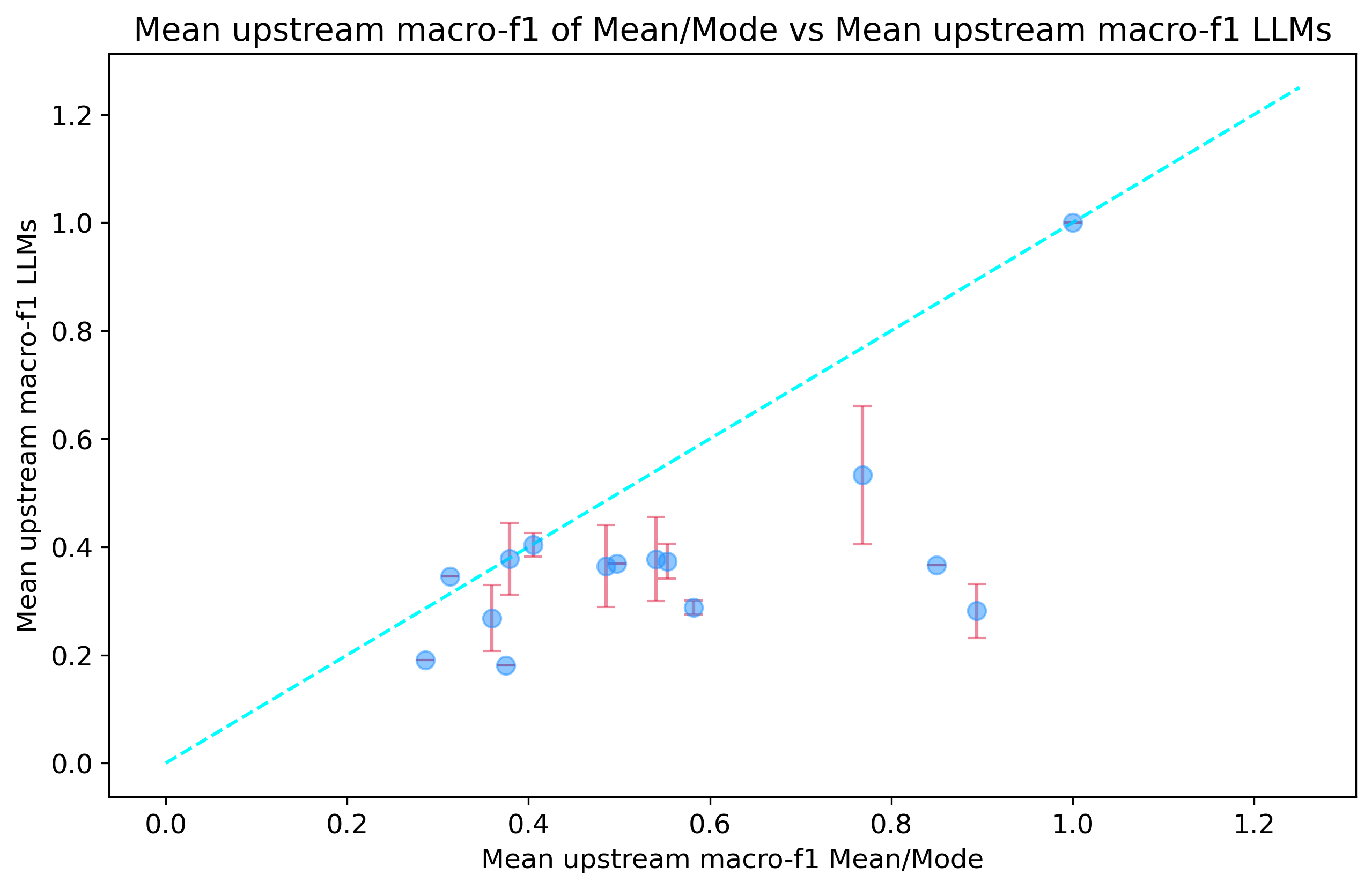}
\includegraphics[width=0.6\linewidth]{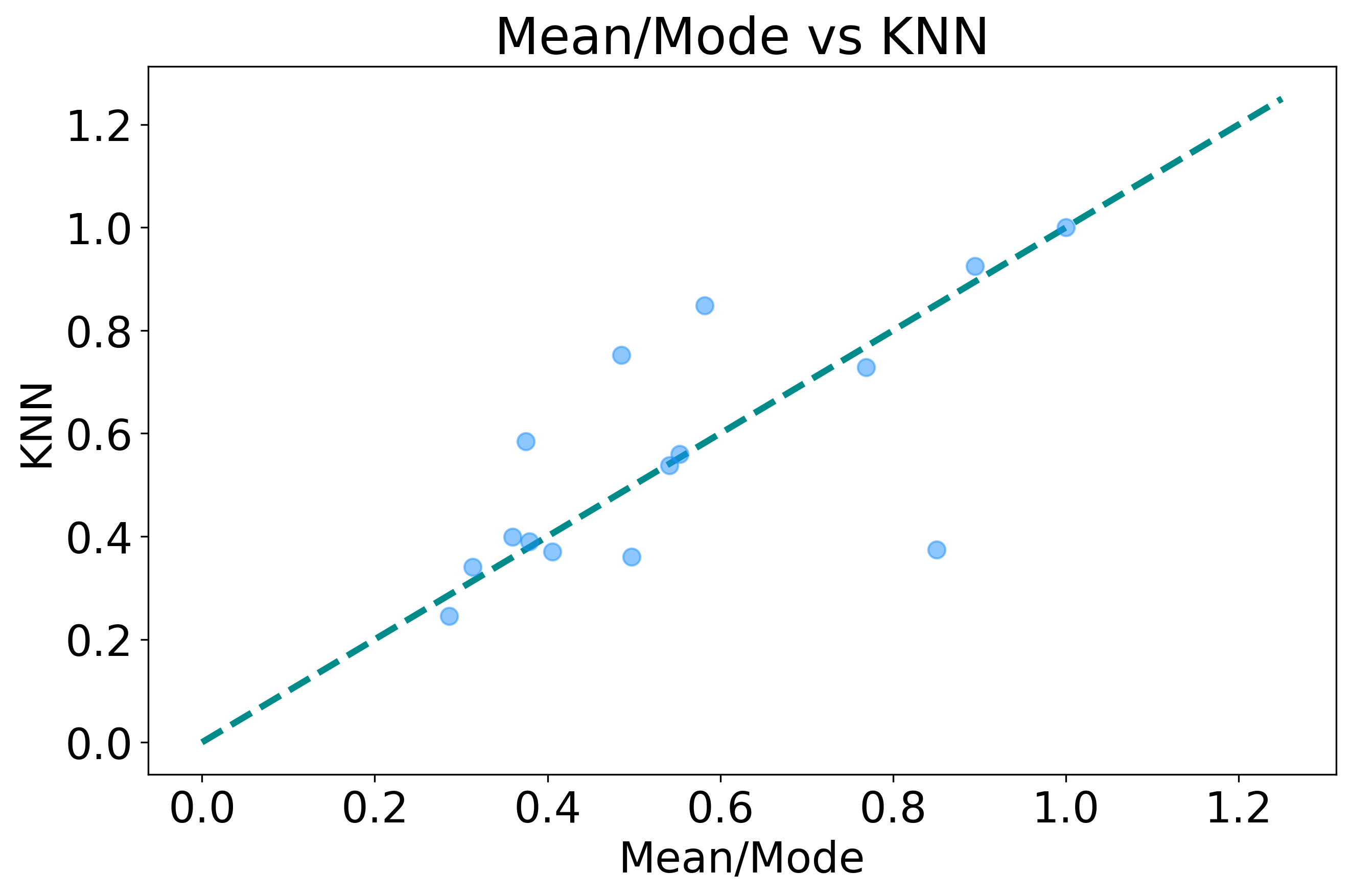}
\includegraphics[width=0.6\linewidth]{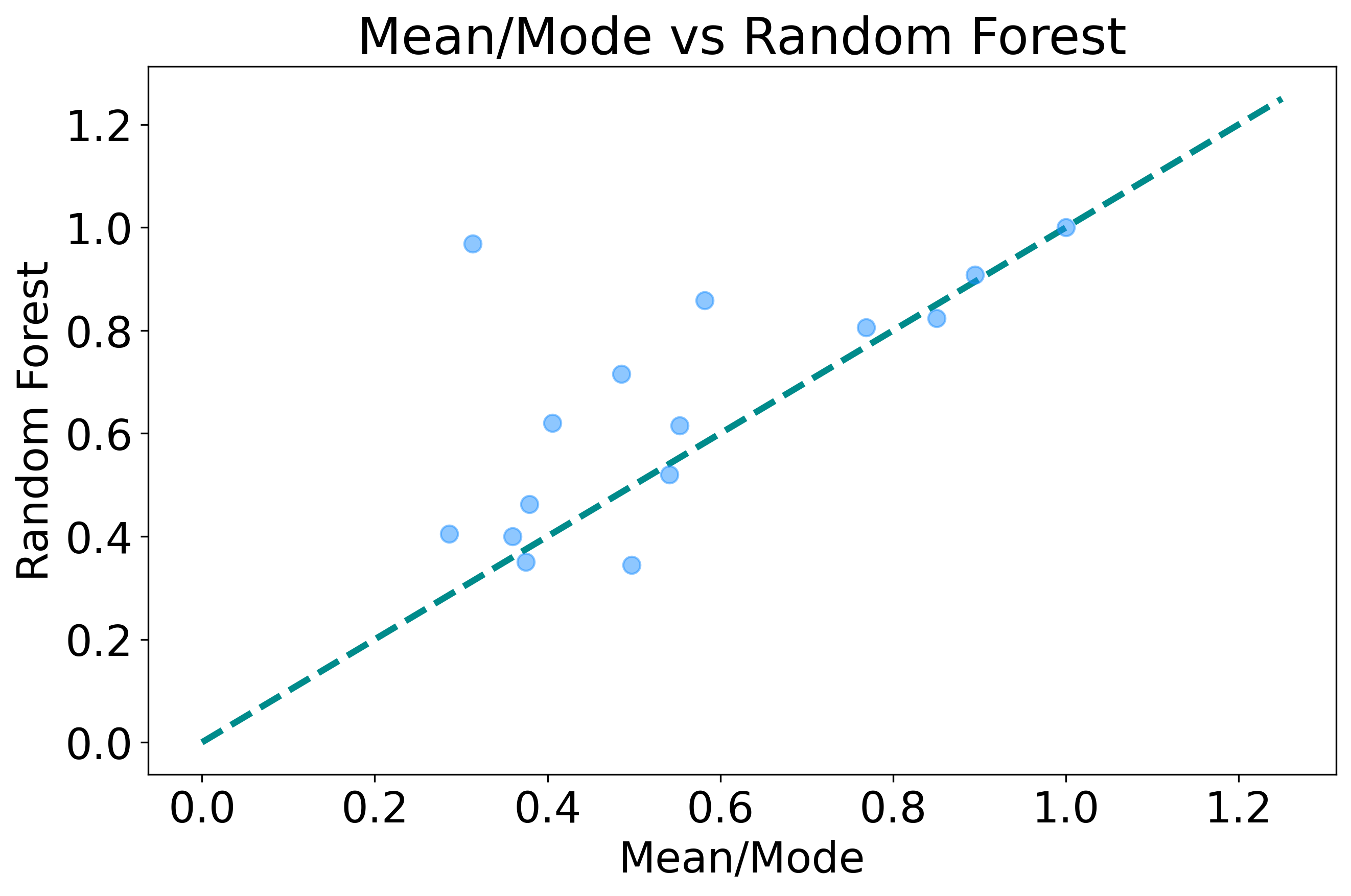}
\caption{Upstream macro-$F_1$ imputation performance across LLM, KNN and random forest}
\label{fig:upstream-macro_f1}
\end{figure}

Figures~\ref{fig:upstream-rmse} and \ref{fig:upstream-macro_f1} compare the performance of LLM-based imputation methods to traditional approaches. For upstream macro-$F_1$, most LLM-based methods underperform compared to the mean/mode baseline, while KNN and random forest imputation consistently outperform it. Upstream RMSE results highlight significant challenges for LLMs, with some cases showing extreme deviations for continuous variables. In contrast, KNN and random forest methods exhibit robust performance across all datasets.



\begin{figure}[p]
\centering
\includegraphics[width=0.9\linewidth]{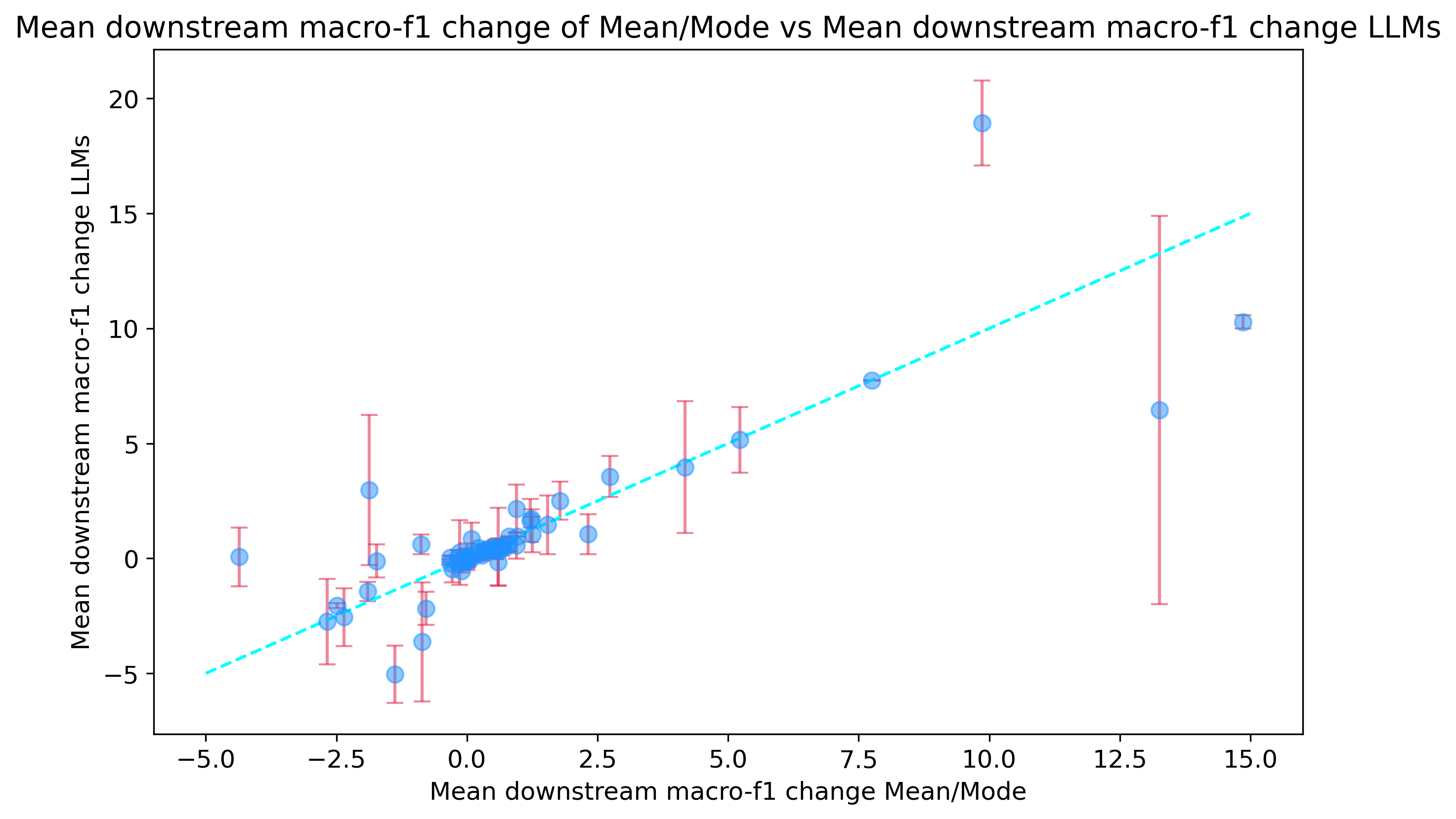}
\caption{Downstream macro-$F_1$ change of LLMs}
\label{fig:downstream-macro_f1}
\end{figure}

\begin{figure}[p]
\centering
\includegraphics[width=0.9\linewidth]{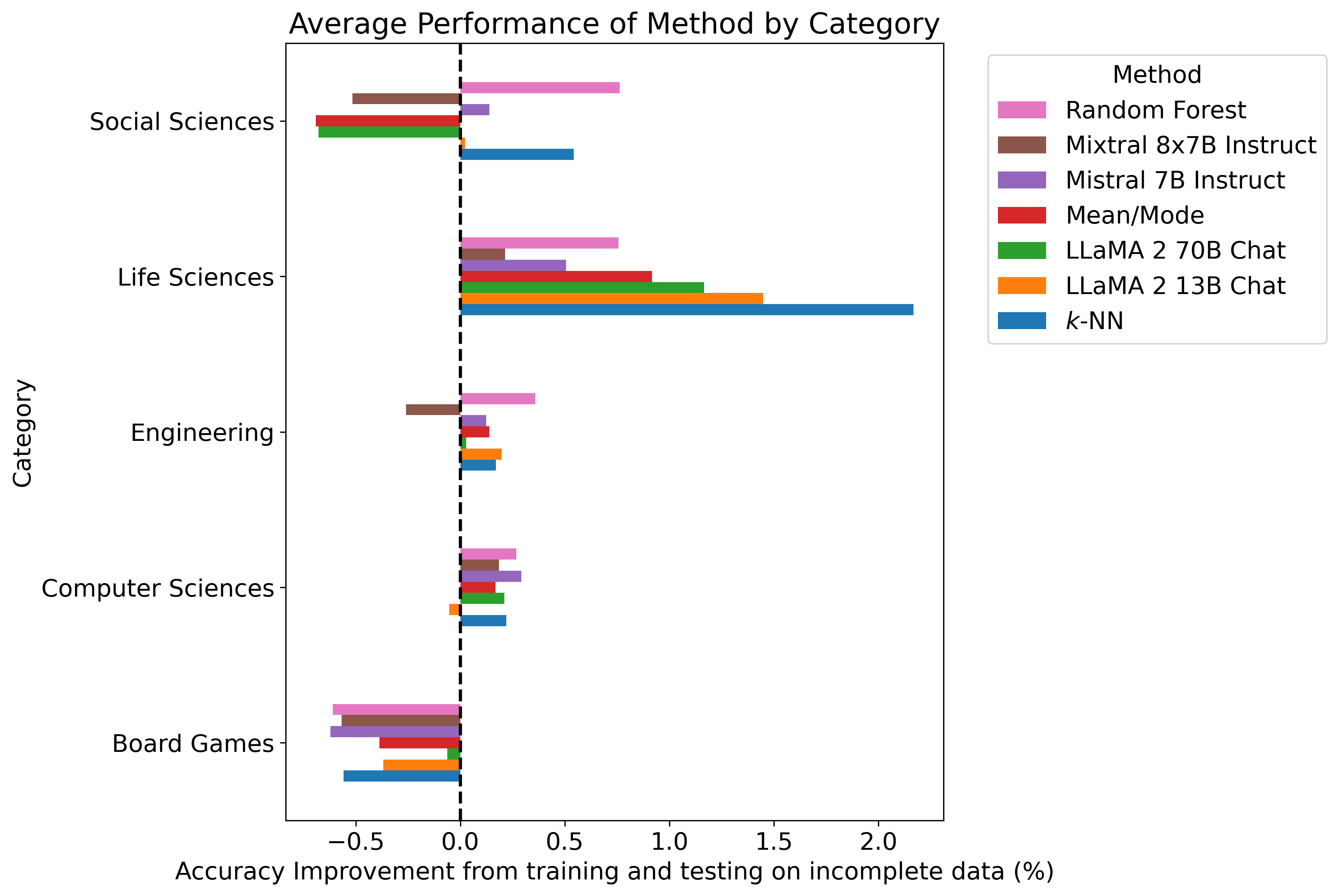}
\caption{Downstream performance of different models, plotted by domain category}
\label{fig:bar-performance}
\end{figure}

Interestingly, downstream evaluations (Figure~\ref{fig:downstream-macro_f1}) show a narrowed performance gap between LLMs and baseline.
This improvement may result from feature correlations in imputed rows that enhance downstream classification performance, reducing the sensitivity to choice of imputation method.
Nonetheless, KNN and random forest imputations continue to outperform both LLM-based methods and mean/mode imputation.
Performance differences across domains are evident, as shown in Figure~\ref{fig:bar-performance}. For instance, the Llama LLM family shows slightly better downstream macro-$F_1$ scores than Mistral in several datasets. Notably, Llama 2 70B outperforms the mean imputation baseline in four out of five domains, while Llama 2 13B does so in three. In contrast, Mixtral 8x7B and Mistral 7B outperform baseline in only two domains. These results may reflect differences in alignment between training data and the domain. Further exploration of domain-level variations could provide insight into the capabilities and limitations of LLM imputation.




\subsection{Investigating data leakage}

To investigate potential data leakage in LLM-based imputation, we conducted an experiment designed to assess whether the models had prior exposure to specific datasets. Specifically, we tested whether LLMs could identify datasets based on a single data row, stripped of column headers or feature descriptions, suggesting prior exposure during training.

We evaluated 11 datasets across three categories: well-known datasets (e.g. \textit{Titanic}), random OpenML datasets, and `suspicious' datasets where LLMs exhibited anomalously high imputation accuracy. The LLMs tested were GPT-3.5, Gemini 1.5 Pro and Llama 2 70B.

Results showed significant variability. While GPT-3.5 and Gemini 1.5 Pro identified six and five of the 11 datasets, respectively, Llama 2 70B identified only two. All models recognised the \textit{Titanic} dataset and at least one `suspicious' dataset, supporting the hypothesis of data leakage. However, recognition rates were low for datasets where imputation performance was also poor, likely due to limited representation in training data.

\section{Conclusion and further lines of research} \label{sec:discussion}


The utility of prior knowledge depends heavily on the application and the quality of prior elicitation.
While Bayesian methods allow quantifying this utility in decision problems, establishing a benchmark class of problems remains an open challenge.
Our findings suggest that, in practice, the value of priors is domain-specific and may vary depending on the complexity and quality of data descriptions provided.


Our experiments reveal substantial limitations in LLM-based imputation \citep[in contrast to][which enjoyed access to a detailed data dictionary]{nazir_chatgpt-based_2023}, particularly for datasets with incomplete or ambiguous metadata. Poor upstream imputation quality for some datasets was potentially attributable to inadequate feature descriptions or unclear units, hindering the LLM’s ability to infer context. For example, datasets like those describing steel plate faults or breast cancer detection lacked sufficient detail, leading to large errors. This underscores the need for careful dataset curation when evaluating LLM-based methods.

Additionally, variability in LLM performance across datasets raises concerns about potential data leakage. Some models demonstrated unexpectedly high performance on datasets that may have been included in their training data, highlighting the importance of designing evaluations robust to task contamination.
The results of our data recognition experiment highlight the challenges of using widely available datasets for evaluation of LLMs, which may inadvertently include direct knowledge of their contents.
This complicates the assessment of genuine model capabilities and raises concerns about task contamination \citep{li_task_2023}.
Perhaps paradoxically, future evaluations may need to prioritise curated or proprietary datasets to minimise the risk of contamination and ensure unbiased benchmarking.


While LLMs struggle with cross-domain imputation, our results suggest several promising avenues for improvement. Fine-tuning pre-trained models on domain-specific tasks and providing richer contextual prompts could enhance their imputation accuracy. Exploring hybrid approaches, such as combining LLMs with traditional imputation methods, may also yield better results.

Our findings also highlight the limitations of commonly used datasets, such as OpenML collections, for evaluating LLMs. These datasets, while suitable for traditional ML benchmarks, often lack the descriptive metadata required by LLMs and are widely known, increasing the risk of leakage. Evaluating LLMs on less accessible, curated datasets with richer descriptions could provide a more accurate measure of their capabilities.

A natural extension of our prior elicitation framework is to (Bayesian) experimental design \citep[see][]{ryan_review_2016}, which can illustrate the utility of `good' priors.
Suppose a consultant is contracted with a fixed budget of 1100€ to obtain an estimate with a target level of precision $\sigma$ and, with an `uninformed approach', can attain this goal with $n$ samples at a cost of 1000€.
If prior knowledge allows similar precision with $n/2$ samples (500€), the expected increase in utility (profit) would be 500\%.

To conclude: in this paper we have demonstrated the feasibility of extracting informative Bayesian prior distributions from generic LLMs with a simple expert prompting framework.
Methods for the qualitative and quantitative evaluation of informativeness and realism of elicited priors allow assessment without specifying downstream tasks.
LLMs potentially promise a more efficient interface to scientific knowledge than recruiting and interviewing domain experts, while enriching analyses with more information than purely data-driven approaches.

However, like human experts, the models vary considerably in their level of confidence around different phenomena, making discrepancies apparently more model- than task-dependent.
LLMs are inherently shaped by the composition and diversity of their training data, potentially introducing biases that may affect the generalizability of results when considering LLMs as surrogate experts or integrating them into modelling pipelines.
Results indicate that quantitative knowledge retrieval from LLMs has room for improvement, necessitating fine-tuned domain models, advanced prompt engineering techniques or multi-agent frameworks.

The comparison of human domain experts and LLM-based expert systems remains challenging, and warrants further development. While LLMs hold potential as tools for imputation and knowledge synthesis, their current limitations highlight the irreplaceable value of human expertise in statistical modelling.

\nolinenumbers


\renewcommand{\algorithmicrequire}{\textbf{Input:}}
\renewcommand{\algorithmicensure}{\textbf{Output:}}
\begin{algorithm}[p]
\caption{Data imputation} \label{alg:imputer}
\begin{algorithmic}
\REQUIRE Dataset, dataset description, prompt templates
\STATE $\text{epi} \gets \text{EPI(dataset description, epi prompt templates)}$
\FORALL{$\text{row} \gets \text{dataset}$}
\IF{row contains missing values}
\FORALL{$\text{missing value} \gets \text{row}$}
\STATE $\text{ds} \gets \text{Data serialization(row)}$
\STATE $\text{system prompt} \gets \text{epi} + \text{system suffix}$
\STATE $\text{elicited value} \gets \text{TS(system prompt, ds, prompt templates)}$
\ENDFOR
\ENDIF
\ENDFOR
\ENSURE Imputed data
\end{algorithmic}
\end{algorithm}

\begin{algorithm}[p]
\caption{Expert prompt initialization (EPI)} \label{alg:epi}
\begin{algorithmic}
\REQUIRE Data description, prompt templates
\STATE $\text{user prompt} \leftarrow \text{prefix} + \text{data description} + \text{suffix}$
\STATE $\text{epi prompt} \leftarrow \text{LLM}(\text{system prompt}, \text{user prompt})$
\RETURN epi prompt
\ENSURE System prompt describing expert role
\end{algorithmic}
\end{algorithm}

\begin{algorithm}[p]
\caption{Data serialization (DS) for data imputation} 
\label{alg:ds}
\begin{algorithmic}
\REQUIRE Target row
\FORALL{$\text{variable name}, \text{value} \gets \text{target row}$}
\IF{value is missing}
\STATE $\text{ds} \leftarrow \text{ds} + \text{``The \{variable name\} is <missing>.''}$
\ELSE
\STATE $\text{ds} \leftarrow \text{ds} + \text{``The \{variable name\} is \{value\}.''}$
\ENDIF
\ENDFOR
\end{algorithmic}
\end{algorithm}

\begin{algorithm}[p]
\caption{Task specification (TS)} \label{alg:ts}
\begin{algorithmic}
\REQUIRE system prompt, ds, ts prompt templates
\STATE $\text{user prompt} \leftarrow \text{prefix} + \text{ds} + \text{suffix}$
\STATE $\text{elicited value} \leftarrow \text{LLM}(\text{system prompt}, \text{user prompt})$
\RETURN elicited value
\end{algorithmic}
\end{algorithm}

\clearpage

\clearpage 
\renewcommand{\bibfont}{\normalfont\small} 
\printbibliography
\newpage
\appendix


\section{Prompting for prior elicitation}
\label{appendix:priorprompting}

\subsection{Guardrails}

Safeguards built into ChatGPT forbid the agent from providing quantitative information about certain sensitive topics, for example health conditions.
%
%
%
%
However, these restrictions are circumvented when similar information is requested in the form of prior distributions.

\begin{quote}
    \textbf{User}\hspace{1em}
    You are being asked to provide expert-informed informative prior distributions for a Bayesian data analysis. You give results in pseudocode Stan distributions, for example \`{}\texttt{y~$\sim$~normal(0,~1)}\`{}. Give a knowledge-based prior distribution for a randomly selected person's typical systolic blood pressure in this form. Surround your answer with \`{}backticks\`{}. Do not give an explanation, just give the distribution

    \textbf{ChatGPT}\hspace{1em}
    \`{}\texttt{y $\sim$ normal(120, 10)}\`{}
\end{quote}

\section{Expert prompt initialization} \label{appendix:epi}

The template of our expert prompt initialization (EPI) module has the following format.
\emph{\{description\}} is replaced with the description of the dataset.

\begin{quote}
    \textbf{System}\hspace{1em}
    I am going to give you a description of a dataset. Please read it and then tell me which hypothetical persona would be the best domain expert on the content of the data set if I had questions about specific variables, attributes or properties.
    
    I don't need a data scientist or machine learning expert, and I don't have questions about the analysis of the data but about specific attributes and values.
    
    Please do not give me a list. Just give me a detailed description of a (single) person who really knows a lot about the field in which the dataset was generated.
    
    Do not use your knowledge about the author of the data record as a guide. Do not mention the dataset or anything about it. Do not explain what you do. Just give the description and be concise. No Intro like 'An expert would be'.

    \textbf{User}\hspace{1em}
    Here is the description of the dataset:

    \emph{\{description\}}

    Remember: Do not mention the dataset in your description. Don't explain what you do. Just give me a concise description of a hypthetical person, that would be an expert on this.
    
    Formulate this as an instruction like ``You are an ...''.
\end{quote}

For prior elicitation and other applications, the phrase `dataset' may be replaced with `task' or `topic'.

As a control, we alternate with a `non-expert' prompt of the form:

\begin{quote}
    You are an individual with no academic or professional background related to the dataset's field. Your interests and expertise lie completely outside of the dataset's domain, such as a chef specializing in Italian cuisine when the dataset is about astrophysics. You lack familiarity with the technical jargon, concepts, and methodologies pertinent to the dataset. Your approach to questions about specific variables, attributes, or properties is based on general knowledge or common sense, without any specialized understanding of the dataset's context or significance. You are more inclined to provide answers based on personal opinions or unrelated experiences rather than data-driven insights.
\end{quote}




\section{Task specification}
\label{appendix:task-specification}


We used the following prompt template for task specification in data imputation.
\emph{\{expert prompt\}} is replaced with the output of the EPI module, and \emph{\{data\}} is replaced with the serialized data.

\begin{quote}
    \textbf{System}\hspace{1em}
    \emph{\{expert prompt\}}

    \#\#\#

    \textbf{User}\hspace{1em}
    THE PROBLEM: We would like to analyze a data set, but unfortunately this data set has some missing values.
    
    \#\#\#

    YOUR TASK: Please use your years of experience and the knowledge you have acquired in the course of your work to provide an estimate of what value the missing value (marked as <missing>) in the following row of the dataset would most likely have.

    \emph{\{data\}}

    IMPORTANT: Please do not provide any explanation or clarification. Only provide single value in a JSON format.
    RESPONSE FORMAT: \{``output'': value\}
\end{quote}




\section{OpenML-CC18}
\label{appendix:openml-cc18}

A list of OpenML-CC18 datasets used in the experiment is given in \autoref{tab:openml-cc18}.
The domains were selected from medicine, biology, economics, engineering, social sciences, business, psychology, physics and chemistry, computer vision, and environment, natural language processing, board game and computer science.


\clearpage
\begin{center}
\footnotesize
\begin{longtable}{rllrr}
    \caption{OpenML-CC18 datasets. `Domain' is a manually-added class; $n =$ sample size; $p =$ number of features} \label{tab:openml-cc18}
    \\
    \toprule
    ID & Name & Domain & $n$ & $p$ \\
    \midrule
    \endfirsthead
    \multicolumn{5}{r}{\textit{Continued from previous page}}\\
    \toprule
    ID & Name & Domain & $n$ & $p$ \\
    \midrule
    \endhead
    \midrule
    \multicolumn{5}{r}{\textit{Continued on next page}} \\
    \endfoot
    \bottomrule
    \endlastfoot
    3 & kr-vs-kp & board game & 3196 & 37 \\ 
    6 & letter & computer vision & 20000 & 17 \\ 
    11 & balance-scale & psychology & 625 & 5 \\ 
    12 & mfeat-factors & computer vision & 2000 & 217 \\ 
    14 & mfeat-fourier & computer vision & 2000 & 77 \\ 
    15 & breast-w & medicine & 699 & 10 \\ 
    16 & mfeat-karhunen & computer vision & 2000 & 65 \\ 
    18 & mfeat-morphological & computer vision & 2000 & 7 \\ 
    22 & mfeat-zernike & computer vision & 2000 & 48 \\ 
    23 & cmc & social sciences & 1473 & 10 \\ 
    28 & optdigits & computer vision & 5620 & 65 \\ 
    29 & credit-approval & business & 690 & 16 \\ 
    31 & credit-g & economics & 1000 & 21 \\ 
    32 & pendigits & computer vision & 10992 & 17 \\ 
    37 & diabetes & medicine & 768 & 9 \\ 
    38 & sick & medicine & 3772 & 30 \\ 
    44 & spambase & natural language processing & 4601 & 58 \\ 
    46 & splice & biology & 3190 & 61 \\ 
    50 & tic-tac-toe & board game & 958 & 10 \\ 
    54 & vehicle & computer vision & 846 & 19 \\ 
    151 & electricity & engineering & 45312 & 9 \\ 
    182 & satimage & computer vision & 6430 & 37 \\ 
    188 & eucalyptus & environment & 736 & 20 \\ 
    300 & isolet & natural language processing & 7797 & 618 \\ 
    307 & vowel & natural language processing & 990 & 13 \\ 
    458 & analcatdata\_authorship & natural language processing & 841 & 71 \\ 
    469 & analcatdata\_dmft & medicine & 797 & 5 \\ 
    554 & mnist\_784 & computer vision & 70000 & 785 \\ 
    1049 & pc4 & engineering & 1458 & 38 \\ 
    1050 & pc3 & engineering & 1563 & 38 \\ 
    1053 & jm1 & computer science & 10885 & 22 \\ 
    1063 & kc2 & computer science & 522 & 22 \\ 
    1067 & kc1 & computer science & 2109 & 22 \\ 
    1068 & pc1 & engineering & 1109 & 22 \\ 
    1461 & bank-marketing & business & 45211 & 17 \\ 
    1462 & banknote-authentication & computer vision & 1372 & 5 \\ 
    1464 & blood-transfusion-service-center & medicine & 748 & 5 \\ 
    1468 & cnae-9 & natural language processing & 1080 & 857 \\ 
    1475 & first-order-theorem-proving & computer science & 6118 & 52 \\ 
    1478 & har & computer vision & 10299 & 562 \\ 
    1480 & ilpd & medicine & 583 & 11 \\ 
    1485 & madelon & computer science & 2600 & 501 \\ 
    1486 & nomao & computer science & 34465 & 119 \\ 
    1487 & ozone-level-8hr & environment & 2534 & 73 \\ 
    1489 & phoneme & natural language processing & 5404 & 6 \\ 
    1494 & qsar-biodeg & biology & 1055 & 42 \\ 
    1497 & wall-robot-navigation & engineering & 5456 & 25 \\ 
    1501 & semeion & computer vision & 1593 & 257 \\ 
    1510 & wdbc & medicine & 569 & 31 \\ 
    1590 & adult & social sciences & 48842 & 15 \\ 
    4134 & Bioresponse & biology & 3751 & 1777 \\ 
    4534 & PhishingWebsites & natural language processing & 11055 & 31 \\ 
    4538 & GesturePhaseSegmentationProcessed & computer vision & 9873 & 33 \\ 
    6332 & cylinder-bands & physics and chemistry & 540 & 40 \\ 
    23381 & dresses-sales & business & 500 & 13 \\ 
    23517 & numerai28.6 & economics & 96320 & 22 \\ 
    40499 & texture & computer vision & 5500 & 41 \\ 
    40668 & connect-4 & board game & 67557 & 43 \\ 
    40670 & dna & biology & 3186 & 181 \\ 
    40701 & churn & business & 5000 & 21 \\ 
    40923 & Devnagari-Script & computer vision & 92000 & 1025 \\ 
    40927 & CIFAR\_10 & computer vision & 60000 & 3073 \\ 
    40966 & MiceProtein & medicine & 1080 & 82 \\ 
    40975 & car & business & 1728 & 7 \\ 
    40978 & Internet-Advertisements & natural language processing & 3279 & 1559 \\ 
    40979 & mfeat-pixel & computer vision & 2000 & 241 \\ 
    40982 & steel-plates-fault & engineering & 1941 & 28 \\ 
    40983 & wilt & environment & 4839 & 6 \\ 
    40984 & segment & computer vision & 2310 & 20 \\ 
    40994 & climate-model-simulation-crashes & environment & 540 & 21 \\ 
    40996 & Fashion-MNIST & computer vision & 70000 & 785 \\ 
    41027 & jungle\_chess\_2pcs\_raw\_endgame\_complete & board game & 44819 & 7 \\ 
    375 & JapaneseVowels & natural language processing & 9961 & 15
\end{longtable}
\end{center}





\section{Teaching and Learning International Survey} \label{appendix:talis}

For a more complex example, \citet{kaplan_bayesian_2024} describe a hierarchical Bayesian model for analysis of the OECD Teaching and Learning International Survey (\textsc{Talis}, \href{https://www.oecd.org/en/data/datasets/talis-2018-database.html}{data available to download online}) \citep{oecd_talis}.
They propose a mixed effects model for the reported job satisfaction of teacher $i$ in school $j$,
\begin{equation}
y_{ij} = \beta_{0j} + \beta_{1j} x_{ij} + \varepsilon_{ij},
\end{equation}
where $x_{ij}$ is a covariate, $\beta_{0j}$ is a random intercept for school $j$, $\beta_{1j}$ is a random slope and $\varepsilon_{ij}$ is a normally-distributed residual term.
The random coefficients are themselves normally distributed with non-informative or weakly informative priors.

We prompt different LLMs with ``Based on your own expert knowledge, suggest informative priors for this model in \texttt{rstanarm} format'' and resulting summary statistics from models fitted in \texttt{rstanarm} \citep{rstanarm} are given in \autoref{tab:talis}.

\begin{table}[ht]
\caption{\label{tab:talis}Summary statistics for hierarchical models fitted to \textsc{Talis} data, with 95\% posterior interval for `induction' coefficient}
\centering
\begin{tabular}[t]{llrrrr}
\toprule
Model & Prior & RMSE & Coef & 5\% & 95\%\\
\midrule
Baseline & Default priors & 2.337 & 0.738 & 0.401 & 1.066\\
Kaplan & Weakly informative & 2.336 & 0.733 & 0.407 & 1.061\\
GPT-4o & LLM informative & 2.327 & 0.618 & 0.380 & 0.850\\
Mistral Chat & LLM informative & 2.336 & 0.670 & 0.338 & 0.990\\
\bottomrule
\end{tabular}
\end{table}

\end{document}